# Forecasting Convective Downburst Potential Over The United States Great Plains

Kenneth L. Pryor

Center for Satellite Applications and Research (NOAA/NESDIS), Camp Springs, MD 20746, USA

**Abstract** A favorable environment for downbursts associated with deep convective storm systems that occur over the central and eastern continental United States includes strong static instability with large amounts of convective available potential energy (CAPE) and the presence of a mid-tropospheric layer of dry (low theta-e) air. A Geostationary Operational Environmental Satellite (GOES) sounder-derived wet microburst severity index (WMSI) was developed and implemented to assess the potential magnitude of convective downbursts, incorporating CAPE as well as the vertical theta-e difference (TeD) between the surface and mid-troposphere to infer the presence of a mid-level dry air layer. However, previous research has identified that over the central United States, especially in the Great Plains region, an environment between that favorable for wet and dry microbursts may exist during the convective season, resulting in the generation of "hybrid" type microbursts. Hybrid microbursts have been found to originate from deep convective storms that generate heavy precipitation, with sub-cloud evaporation of precipitation a significant factor in downdraft acceleration. Accordingly, a new GOES sounder derived product is under development that is designed to assess the potential for convective downbursts that develop in an intermediate environment between a "wet" type, associated with heavy precipitation, and a "dry" type associated with convection in which very little to no precipitation is observed at the surface. The GOES Hybrid Microburst Index (HMI) algorithm is designed to infer the presence of a convective boundary layer by incorporating the sub-cloud temperature lapse rate (between the 670 and 850 millibar (mb) levels) as well as the dew point depression difference between the typical level of a convective cloud base (670 mb) and the sub-cloud layer (850 mb). Used in conjunction with the GOES WMSI product, the HMI product is intended to indicate the potential magnitude of convective downbursts associated with intermediate type thermodynamic environments over the Great Plains of the United States. In addition, it has been found that the GOES HMI product can be an effective indicator of the presence of the dryline, which can serve as an initiating mechanism for deep convective storm activity. Preliminary validation, using surface observation data from the Oklahoma Mesonet for the period of 1 June to 31 August 2005, has shown favorable results for the coordinated use of the GOES HMI and WMSI. A statistically significant correlation of 0.69 has been found between GOES WMSI values and the magnitude of convective wind gusts for HMI values greater than 16 (considered to be a significant risk). This result highlights the importance of both sub-cloud evaporational cooling as well as static instability in the generation of convective downbursts in an environment typical of the Southern Plains region of the United States. This paper will outline the development of the HMI algorithm and provide examples of the new HMI product, in which index values at each sounding retrieval location are plotted on GOES water vapor imagery. Validation data for the 2005 convective season will be presented. Case studies will then be presented that demonstrate the performance of the coordinated use of the WMSI and HMI during convective events over the Southern Plains. Finally, a modification to convective wind gust prediction with the coordinated use of the WMSI and HMI will be presented.

**Introduction**
A favorable environment for downbursts associated with deep convective storm systems that occur over the central and eastern continental United States includes strong static instability with large amounts of convective available potential energy (CAPE) and the presence of a mid-tropospheric layer of dry (low theta-e) air. CAPE has an important role in precipitation formation due to the strong dependence of updraft strength and resultant precipitation content on positive buoyant energy. Also, mid-tropospheric dry air, laterally entrained into a convective storm cell during downdraft initiation, is instrumental in the increase of negative buoyancy due to evaporational cooling. A Geostationary Operational Environmental Satellite

(GOES) sounder-derived wet microburst severity index (WMSI) (Pryor and Ellrod 2004) was developed and implemented to assess the potential magnitude of convective downbursts, incorporating CAPE as well as the vertical theta-e difference (TeD) between the surface and mid-troposphere to infer the presence of a dry air layer. However, previous research (Fujita 1985, Ellrod 1989) has identified that over the central United States, especially in the Great Plains region, an environment between that favorable for wet microbursts (Atkins and Wakimoto 1991) and dry microbursts (Wakimoto 1985) may exist during the convective season, resulting in the generation of "hybrid" type microbursts. Hybrid microbursts have been found to originate from deep convective storms that generate heavy precipitation, with sub-cloud evaporation of precipitation a significant factor in downdraft acceleration. This intermediate type environment, as described by Caracena et al. (2005), is characterized by conditions favorable for both wet and dry microbursts:

1. Significant CAPE.
2. A deep, dry adiabatic lapse rate layer below the cloud base, which is typically near the 700 mb level.
3. A dry (low theta-e) layer overlying a moist midtropospheric layer.

Accordingly, a new GOES sounder derived product is under development that is designed to indicate the potential for convective downbursts that develop in an intermediate environment between a "wet" type, associated with heavy precipitation, and a "dry" type associated with convection in which very little to no precipitation is observed at the surface. The GOES Hybrid Microburst Index (HMI) algorithm is designed to infer the presence of a convective boundary layer (CBL) by incorporating the sub-cloud temperature lapse rate (between the 670 and 850 millibar (mb) levels) as well as the dew point depression difference between the typical level of a convective cloud base (670 mb) and the sub-cloud layer (850 mb). In a typical dry microburst thermodynamic environment, Wakimoto (1985) identified a convective cloud base height near the 500 mb level. In contrast, Atkins and Wakimoto (1991) identified a typical cloud base height in a pure wet microburst environment near 850 mb. Thus, an intermediate cloud base height of 670 mb was selected for a hypothetical hybrid microburst environment. This selection agrees well with the mean level of free convection (LFC) of 670 mb computed from the inspection of twenty GOES proximity soundings corresponding to downburst events that occurred in Oklahoma between 1 June and 31 July 2005. In a free convective thermodynamic environment (i.e. no convective inhibition (CIN)), the mean LFC of 670 mb can be considered to represent the upper limit for convective cloud base heights that occur in an environment favorable for hybrid microbursts. CAPE, as well as the presence of a mid-tropospheric dry air layer, are already accounted for in the WMSI algorithm. Thus, the Hybrid Microburst Index, intended to serve as a supplemental index to the WMSI, is defined as

$$HMI = G + (T - T_d)_{850} - (T - T_d)_{670} \quad (1)$$

where G is the lapse rate in degrees Celsius (C) per kilometer from the 850 to the 670 mb level, T is temperature in degrees Celsius, and $T_d$ is the dewpoint temperature (C). Inspection of representative proximity soundings revealed that a large HMI value results from a sub-cloud lapse rate that is nearly adiabatic, typically associated with a large lifting condensation level (LCL) and LFC, and a large difference in dew point depression between the approximate level of the convective cloud base (near 670 mb) and the sub-cloud dry air layer (near 850 mb). Figure 1 illustrates the diurnal tendency of HMI values over the High Plains. Note the gradual increase in HMI values over the Oklahoma Panhandle between 1700 and 2000 UTC (1200 and 1500 CDT) 11 September 2005. This increasing trend in HMI values is most likely associated with surface heating and the resultant deepening of the boundary layer between midday and mid-afternoon.

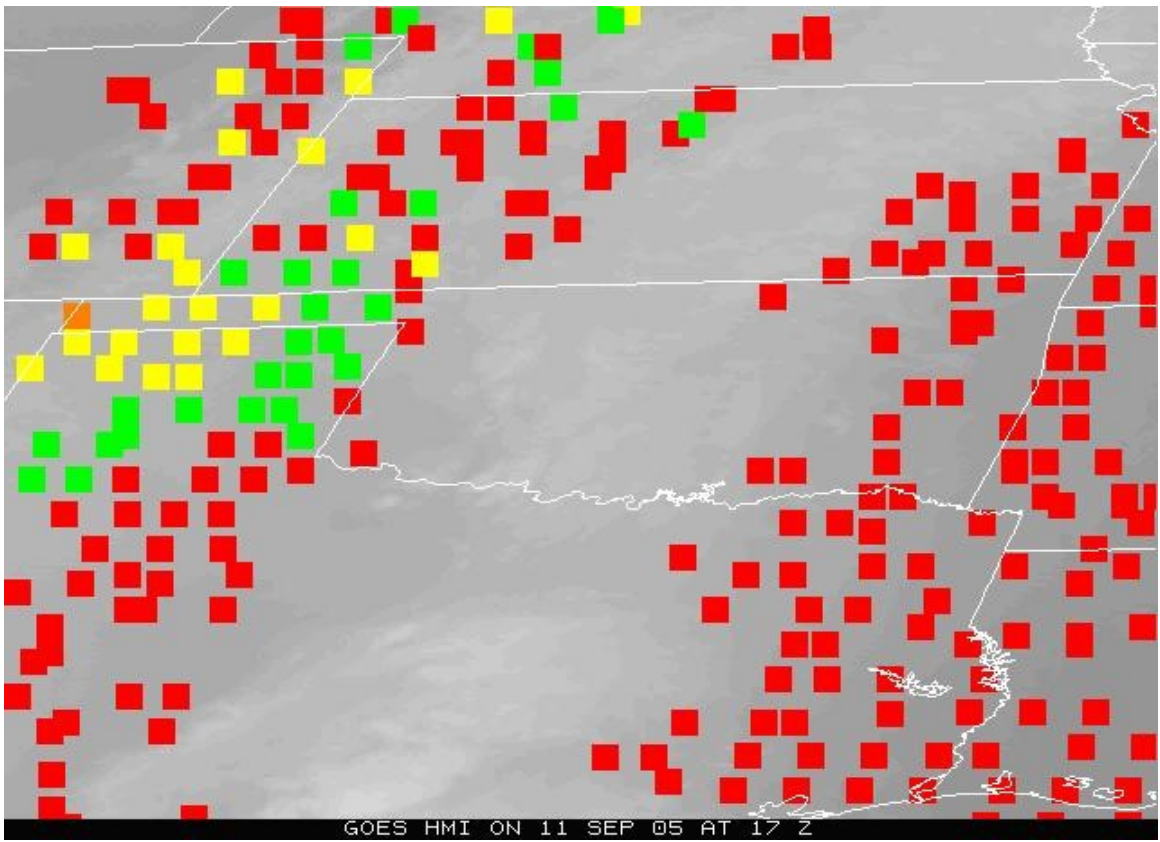

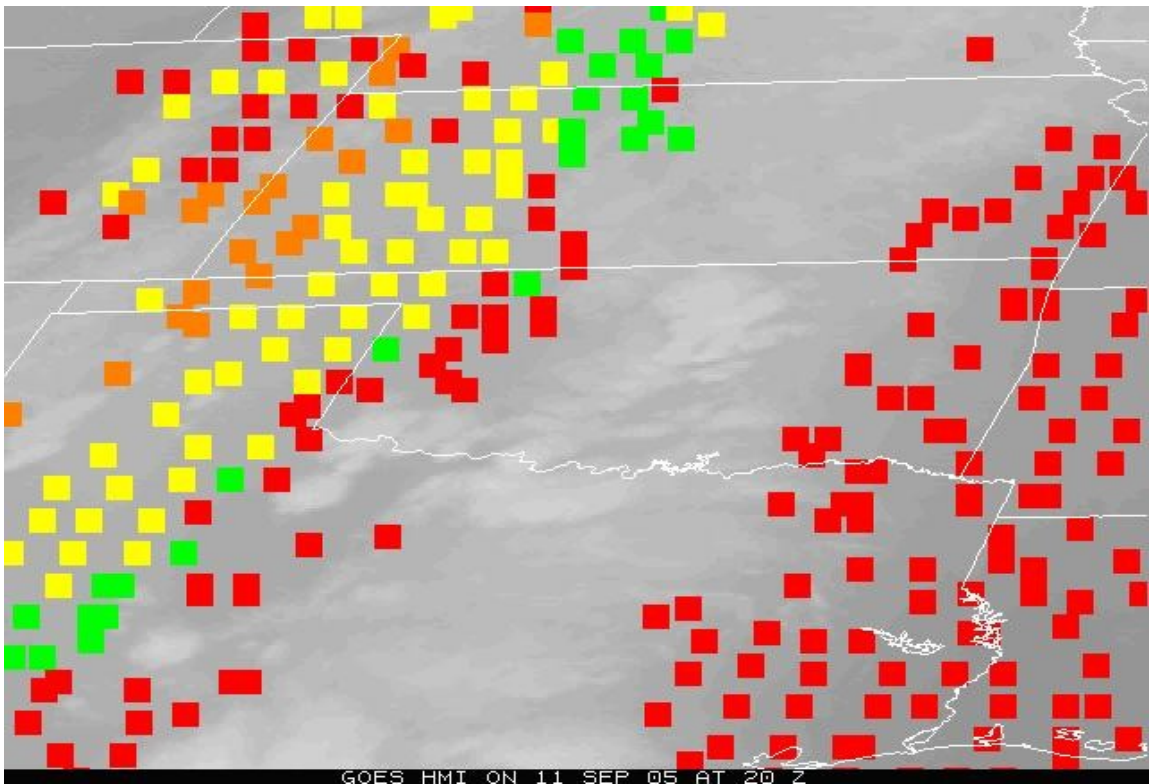

Figure 1. GOES HMI images on 11 September 2005.

---

In a thermodynamic environment favorable for hybrid microbursts, a typical sounding, as displayed in Figure 2 will exhibit an "inverted-v" or "hourglass" profile with a large positive area (CAPE) and a well-defined mid-tropospheric dry air layer.

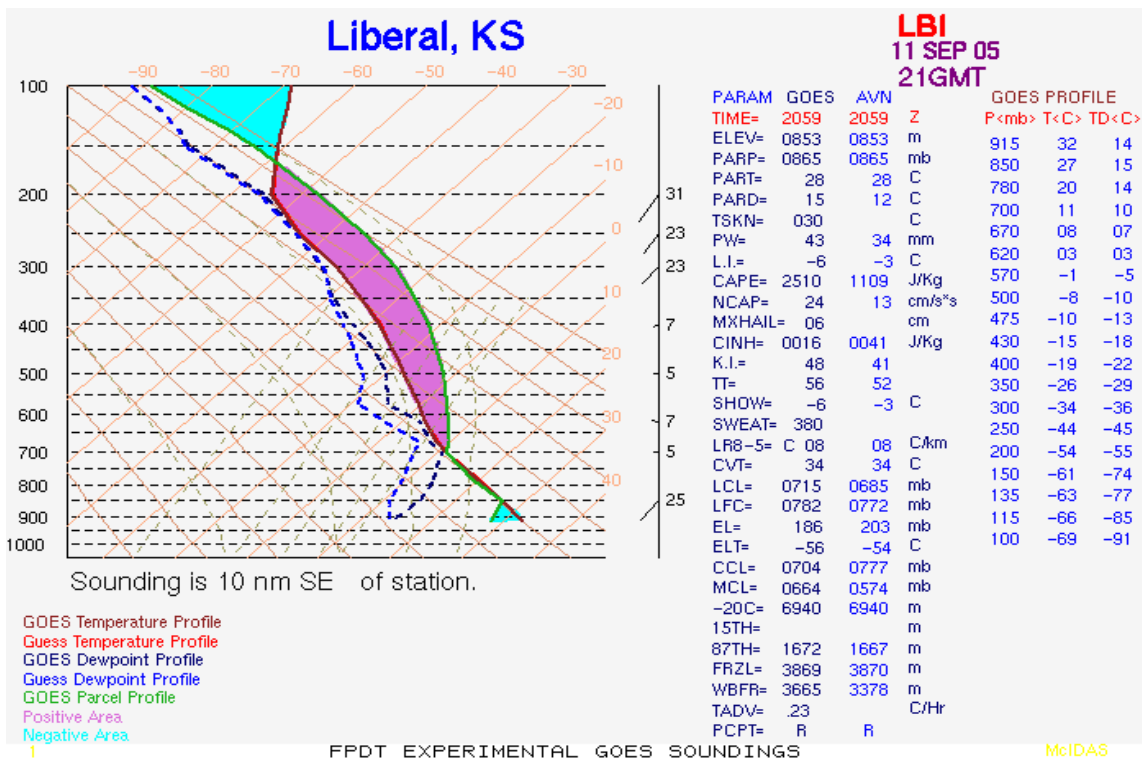

Figure 2. GOES sounding at 2100 UTC 11 September 2005 at Liberal, Kansas.

---

A climatology of severe storm environmental parameters (Nair et al. 2002) has found that a deeper convective mixed layer, as represented by large LFCs and LCLs, predominates in the warm season over the southern Plains. The presence of a deep, dry sub-cloud (mixed) layer will enhance evaporational cooling and downdraft intensification as precipitation falls below the convective storm cloud base. In fact, Nair et al. (2002) have found that moderately high LFCs, that coexist with large CAPE over the Great Plains, are associated with an observed maximum in severe convective storm occurrence. Used in conjunction with the GOES WMSI product, the HMI product is intended for short-term prediction of the magnitude of convective downbursts associated with intermediate type thermodynamic environments over the Great Plains of the United States.

In addition, the thermodynamic structure of the lower and middle troposphere that results in large HMI values signifies the presence of an enhanced convective mixed layer, typically found along the dryline zone over the southern Plains. The dryline is defined as a narrow zone of extremely sharp moisture gradient that separates moist air originating over the Gulf of Mexico from dry air originating from the semi-arid high plateau regions of Mexico and the southwestern United States (Schaefer 1986). Through the dependence of CAPE on low-level moisture, a sharp gradient of WMSI risk values could serve to locate the dryline by inferring the presence of a virtual potential temperature gradient. Stull (1988) has identified that the characteristics of a deep convective mixed layer are caused by a combination of buoyant heat flux, due to strong solar heating of the surface, and wind shear. These conditions are typically found along the dryline. Also, Ziegler and Hane (1993) found the presence of a deeper, well-mixed CBL along and in close proximity to the dryline. In addition, increased vertical circulation, resulting from the sharp temperature and moisture gradients along the dryline zone, is believed to be responsible for enhanced mixing and the subsequent deepening of the CBL and hence, the sub-cloud layer. Thus, it is speculated that the presence of the dryline can establish a thermodynamic setting favorable for hybrid microburst generation by increasing the role of sub-cloud evaporational cooling in the process of downdraft acceleration.

The GOES HMI product appears similarly to the GOES Dry Microburst Index (DMI) product, with risk values plotted over a water vapor satellite image. A sample real-time image is available via FTP:

ftp://ftp.orbit.nesdis.noaa.gov/pub/smcd/opdb/hmi/HMI.GIF

The downburst risk associated with each range of risk values is listed in the table below:

| Table 1. Downburst risk corresponding to HMI values | | | |
|---|---|---|---|
| **HMI** | **Box Color** | **Downburst risk** | |
| < 8 | **Red** | Downbursts Unlikely | |
| > or =8 | **Green** | Downbursts Likely | |
| > or =16 | **Yellow** | Downbursts Likely | |
| > 24 | **Orange** | High Risk of Downbursts | |

**Methodology and Preliminary Validation**

Data from the GOES HMI and WMSI was collected over Oklahoma from 1 June to 31 August 2005 and validated against conventional surface data. The State of Oklahoma was chosen as a study region due to the wealth of surface observation data provided by the Oklahoma Mesonet (Brock et al. 1995), a thermodynamic environment typical of the southern Plains region during the warm season, and its proximity to the dryline. Atkins and Wakimoto (1991) discussed the effectiveness of using mesonet observation data in the verification of the occurrence of downbursts. Validation was conducted in the manner described by Pryor and Ellrod (2004). In addition, GOES sounding profile data, most representative of the preconvective environment, was collected for each downburst event, if available. Correlation between GOES WMSI values and observed surface wind gust velocities, associated with HMI values in each category listed in Table 1, was computed for the period. Hypothesis testing was conducted to determine the statistical significance of linear relationships between observed downburst wind gust magnitude and WMSI values for the HMI value categories presented in Table 1.

The purpose of the validation is to compare the performance of the microburst products for convective downburst events that occur in an intermediate thermodynamic environment characterized by a combination strong static instability and a relatively deep convective mixed layer. A statistically significant correlation of 0.55 was found between GOES WMSI values and the magnitude of convective wind gusts for 72 hybrid microburst events that occurred during the validation period. Partitioning the downburst events by HMI categories resulted in a much stronger correlation of 0.69 for downburst events (N=31) associated with an HMI value greater than 16 (considered to be a significant risk). This result highlights the importance of both sub-cloud evaporational cooling as well as static instability in the generation of convective downbursts in an environment typical of the Southern Plains region of the United States during the warm season. Compared to a correlation between WMSI values and observed wind gusts of 0.66 found over the central and eastern United States (Pryor and Ellrod 2005), this result demonstrates the more robust statistical relationship that can be derived when reducing the validation region of interest from a national scale (i.e. central and eastern U.S.) to a regional scale (i.e. Oklahoma) and conducting validation for a specific thermodyanic environment (i.e. hybrid microbursts).

**Case Study: 11 September 2005 Beaver Microburst**

During the afternoon of 11 September 2005, intense convective storms developed along the dryline over the Oklahoma Panhandle. The presence of the dryline was indicated by a strong virtual potential temperature gradient (not shown) extending from southeastern Colorado to central Oklahoma. Ziegler and Hane (1993), in their observational analysis of a dryline event in western Oklahoma, found a horizontal gradient of virtual potential temperature along and to the east of the dryline. Over the eastern panhandle and northwestern Oklahoma, the pre-convective environment was characterized by moderate static instability and the presence of a deep CBL, resulting from strong solar heating. A comparison of the 1400 and 2100 UTC GOES sounding profiles at Liberal, Kansas (Figures 3 and 1) and a meteogram from Beaver, Oklahoma (Figure 4) display the evolution of the CBL between mid-morning (1400 UTC) through late afternoon (2100 UTC).

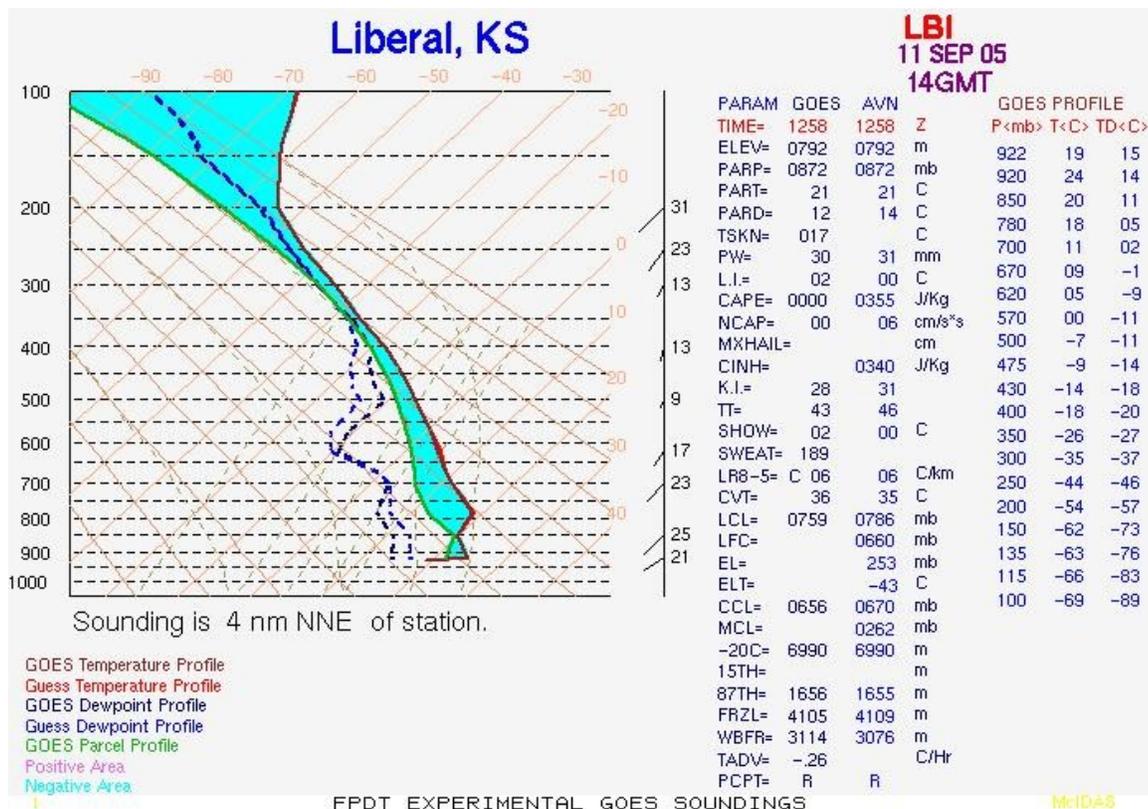

Figure 3. GOES sounding on 11 September 2005 at Liberal, Kansas.

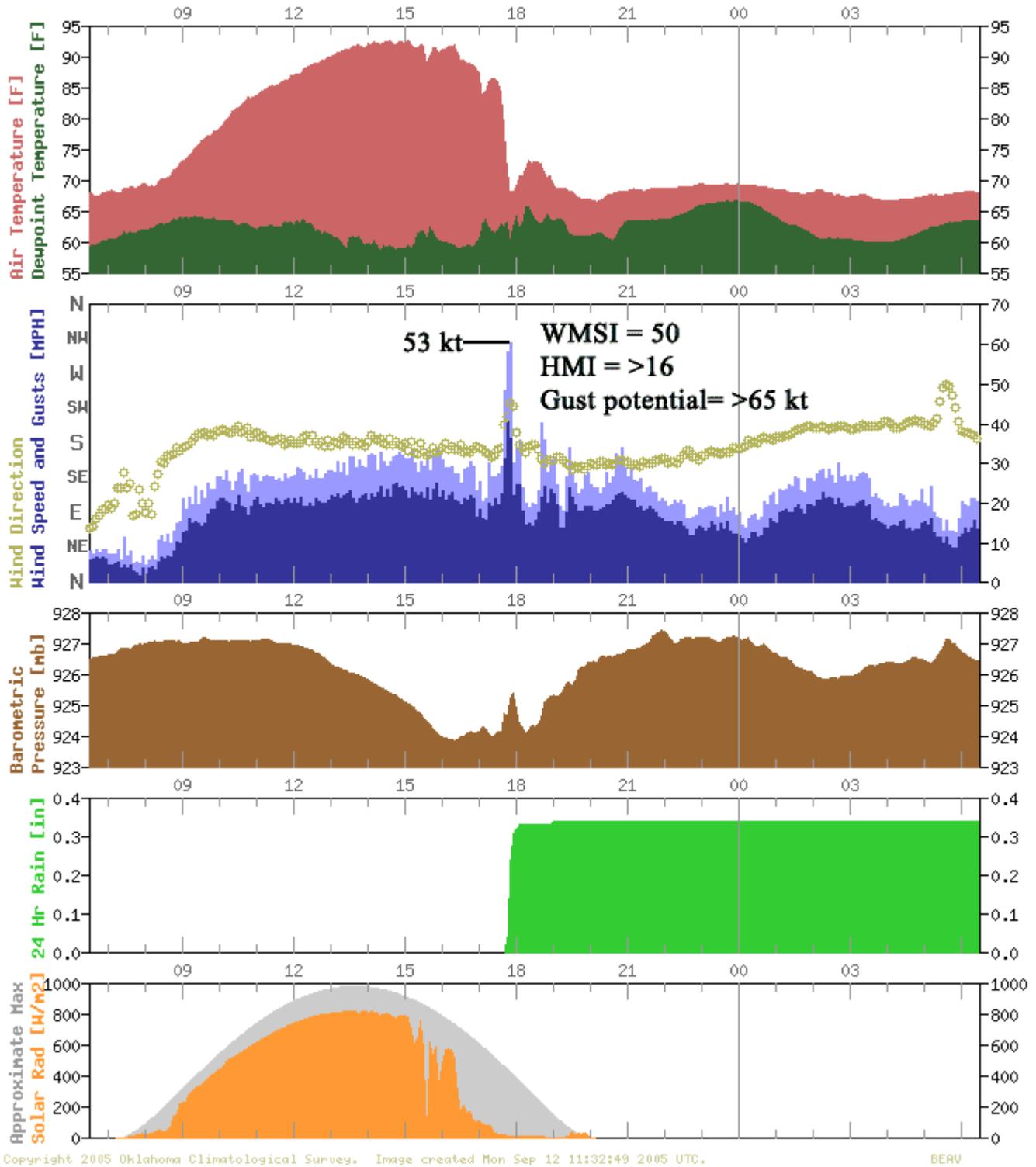

Figure 4. Meteogram from the Beaver, Oklahoma mesonet station (courtesy of the Oklahoma Climatological Survey).

---

Gusty surface winds were associated with the development of the turbulent mixed layer. Note the development of an "inverted-v" profile between 1400 and 2100 UTC as well as a large surface dew point depression (34F or 19C) at Beaver, corresponding to a gradual increase in HMI values over the Oklahoma Panhandle. Stull (1988) noted that a large surface dewpoint depression (> 17C) is associated with a well-developed CBL. By mid-afternoon (2000 UTC), as shown in Figures 1 and 5, HMI values over the eastern panhandle were in excess of 16 with corresponding WMSI values near 50, indicating the presence of an intermediate thermodynamic environment favorable for hybrid microbursts.

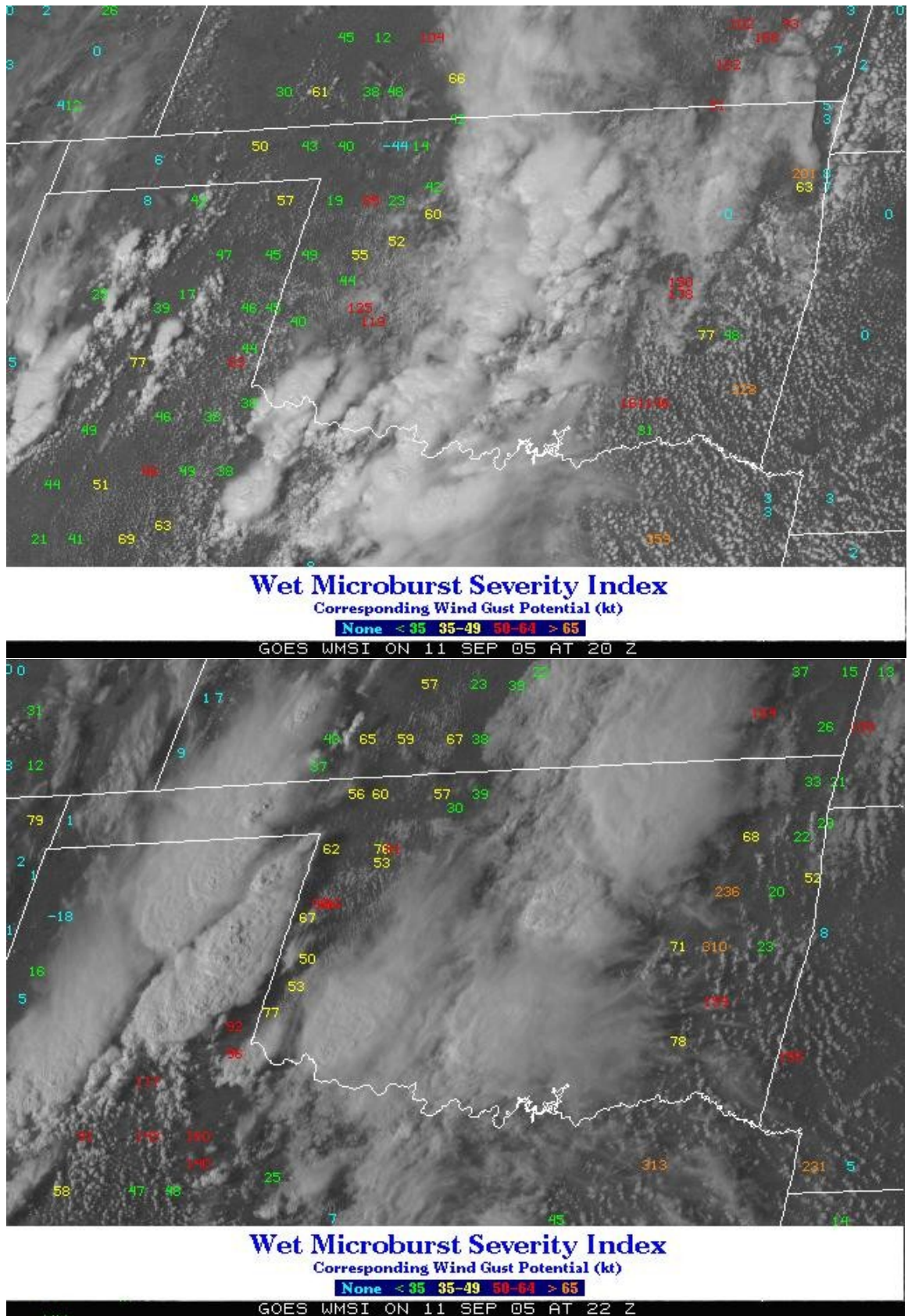

Figure 5. GOES WMSI images on 11 September 2005.

Between 2100 and 2300 UTC an area of convective storms developed over the eastern Oklahoma Panhandle along the dryline, as indicated by both Figure 5 and Figure 6, radar reflectivity imagery from Amarillo, Texas NEXRAD (KAMA).

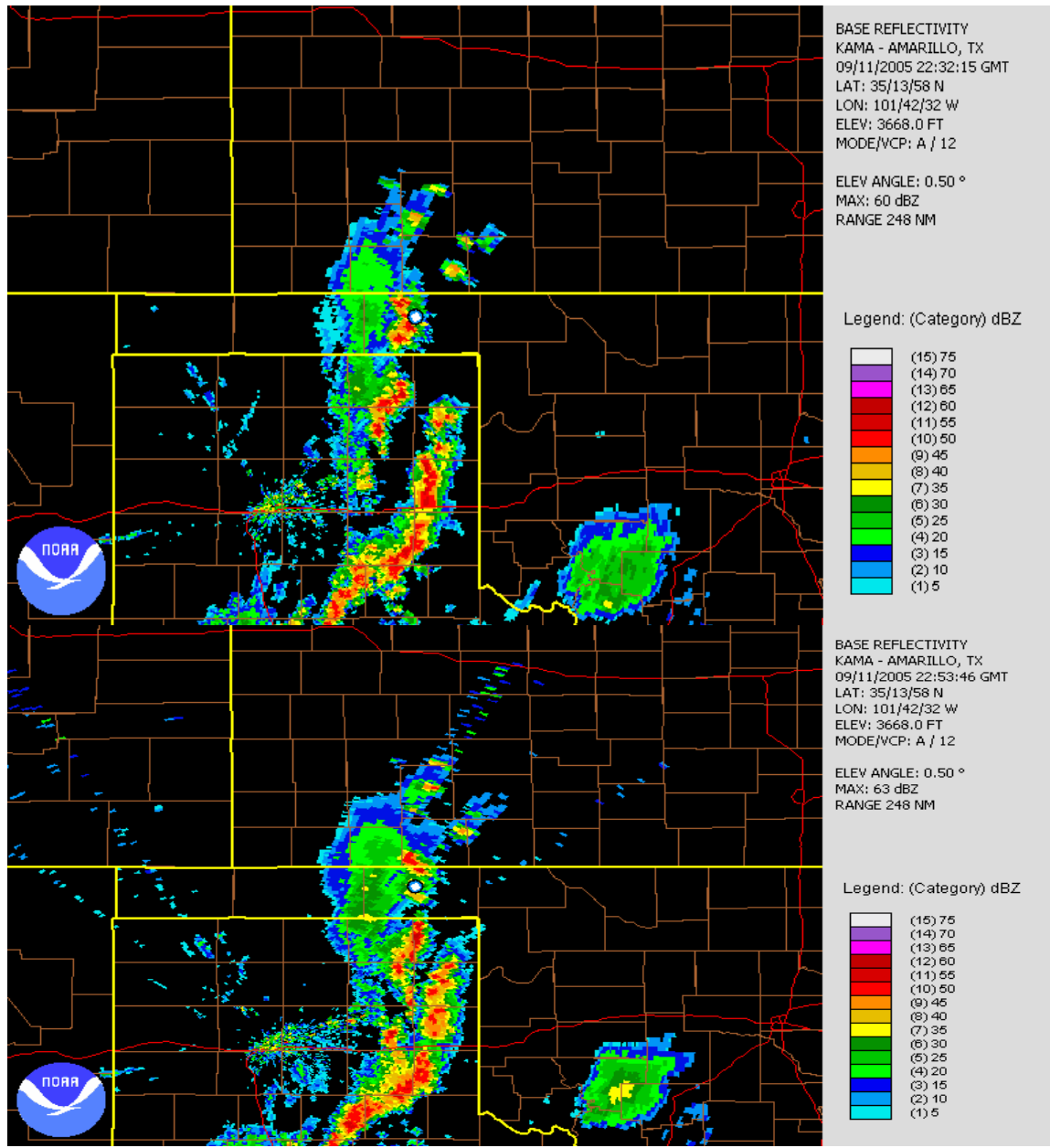

Figure 6. Amarillo, Texas NEXRAD radar reflectivity imagery at 2232 UTC (top) and 2253 UTC (bottom) 11 September 2005. Location of Beaver mesonet station in radar images is indicated by a blue and white marker.

---

By 2232 UTC, the convective storm cluster merged to form a bow echo (Przybylinski 1995) over Beaver County. A severe convective wind gust of 53 knots associated with a microburst was observed at Beaver at 2255 UTC as the bow echo continued to propagate northeastward into Kansas. This case demonstrated the increased role of sub-cloud evaporational cooling in downdraft instability and the subsequent generation of a microburst in an environment of marginal to moderate static instability. As discussed earlier, the dryline zone has been identified as a region of enhanced mixing and vertical circulation. As expected during the warm season, this case emphasizes the effects of bouyancy and stability in convective wind gust magnitude. It is also demonstrated that the presence of the dryline can establish a favorable thermodynamic structure for hybrid microburst generation.

**Conclusions**

The GOES WMSI product was developed to parameterize and approximate the cloud physical and thermodynamic properties (i.e. CAPE, TeD) associated with downbursts that occur over the eastern United States. The HMI product is thus intended to serve as a supplement to the WMSI by inferring the thermodynamic structure of the boundary layer, especially over the Great Plains. Based on previous research and the validation data presented above, the following modification to convective wind gust prediction using the GOES WMSI is proposed:

| Table 2. Convective Wind Gust Prediction Matrix | | | |
|---|---|---|---|
| HMI | WMSI | Wind Gusts (kt) | |
| <8 | 10 - 49 | < 35 | |
|  | 50 - 79 | 35-49 | |
|  | > 80 | >50 | |
| > or =8 | 10 - 30 | < 35 | |
|  | 30 - 49 | 35-49 | |
|  | 50 - 79 | >50 | |
|  | > 80 | > 50 | |
| > or =16 | 10 - 30 | 35 - 49 | |
|  | 30 - 49 | >50 | |
|  | 50 - 79 | >65 | |
|  | > 80 | >65 | |

The data in the table above exemplifies the importance of sub-cloud evaporational cooling of precipitation in the magnitude of convective downbursts. It is apparent that convective storms that develop in an environment of marginal instability (i.e WMSI of 10 to 50) are capable of producing severe downbursts if HMI values are large (> 16).

**Acknowledgements** The author thanks Mr. Derek Arndt (Oklahoma Climatological Survey) and the Oklahoma Mesonet for the surface weather observation data used in this research effort.